\documentstyle[12pt]{article}
\def\ksi{\xi}
\def\be{\begin{equation}}
\def\ee{\end{equation}}
\centerline{\Large Space-Times Admitting Isolated Horizons}

\bigskip

\centerline{\large Jerzy Lewandowski}
\centerline{\it Instytut Fizyki Teoretycznej, Uniwersytet 
Warszawski,} 
\centerline{\it ul. Ho\.za 69, 00-681 Warszawa, Poland, lewand@fuw.edu.pl}

\begin{document}

\begin{abstract} 

We characterize a general solution to the vacuum Einstein equations
which admits  isolated horizons.  We show it is a non-linear
superposition -- in  precise sense -- of the
Schwarzschild metric with a certain free data set propagating
tangentially to the horizon. This proves Ashtekar's conjecture about
the structure of spacetime near the isolated horizon.  The same
superposition method applied to the Kerr metric gives another class of
vacuum solutions admitting isolated horizons. More generally, a vacuum
spacetime admitting any null, non expanding, shear free surface is
characterized.  The results are applied to show that, generically, the
non-rotating isolated horizon does not admit a Killing vector field
and a spacetime is not spherically symmetric near a symmetric horizon.
\end{abstract} 
\newpage

The quantum geometry considerations applied to  black
hole entropy \cite{qentropy} led Ashtekar et al
to a new approach to black hole
mechanics. The idea is to consider a null surface which
locally has the properties of the Schwarzschild horizon, but is not
necessarily infinitely extendible, so the spacetime metric in a
neighborhood is not necessarily that of Schwarzschild. Such a surface
was called a non rotating isolated horizon (NRIH).  The number of
degrees of freedom describing a spacetime admitting a  NRIH
is much larger than that describing a static black hole (see below). 
In a series of works the laws of the black hole thermodynamics and
mechanics were extended to this case \cite{phase,mech}.

In this letter we completely characterize  a general
solution to the Einstein vacuum equations which admits an isolated
horizon and, in particular,  a NRIH. For that purpose, we
use Friedrich's characteristic Cauchy problem defined on null surfaces
\cite{Friedrich} (The idea of constructing solutions to the Einstein's
equations starting with data defined on a null surface was first
formulated by Newman \cite{Newman}). The null Cauchy problem
formulation gives rise to our superposition method: Given a local
solution to the Einstein vacuum equations and the data it defines on a
null surface, a new solution can be constructed from the null surface
data and certain new data freely defined on a transversal null
surface.  We show that a general solution which admits a NRIH is given
by the superposition of the data defined by the Schwarzschild metric
on the horizon and the data defined freely on a transversal null
surface.  This result is then applied to prove that a
generic NRIH does not admit a Killing vector field.  Even
though there are vector fields defined on the horizon which Lie
annihilate the metric tensor \cite{Racz}, none of them, generically,
can be extended to a neighborhood.  The statement concerns the null
vector fields as well as the space like vectors generating symmetries
of the internal geometry induced on the 2 dimensional cross sections
of the NRIH.

We also characterize a general solution to the Einstein vacuum
equations which admits a null non expanding surface.  An interesting
subclass of space-times is obtained by superposing, in our sense, the
data defined by the Kerr metric on its horizon with the data freely
defined on a transversal null surface. By analogy to the non rotating
case, the resulting null surface equipped with the data corresponding
to those of the Kerr metric may be thought of as a rotating isolated
horizon.

Another way to extend our results is to admit matter fields in
spacetime.  In particular the Maxwell field fits the null surfaces
formulations of the Cauchy problem very well.  We 
use Newman-Penrose spin connection and curvature coefficients in
the notation of \cite{exact}.  All our considerations and results will
be {\it local} in the following sense: Given two null 3-surfaces $N_0$
and $N_1$ intersecting in the their future, by {\it locally} we
mean `in the past part of a suitable neighborhood of a $N_0\cap N_1$
bounded by the incoming parts of the surfaces'.

\bigskip

{\bf Isolated horizons: definitions.}
Consider a null 3-submanifold $N_0$ of a 4-dimensional spacetime $M$ 
diffeomorphic to 
\begin{equation}\label{times}
S_2 \times [v_0,v_1],
\end{equation}
where 2-spheres $S_2$  can be identified with
 space-like cross sections and the intervals $[v_0, v_1]$
lie along the null generators  of $N_0$. We say that $N_0$ is an
isolated horizon if the intrinsic, degenerate  metric
tensor  induced in $N_0$ is annihilated by the Lie
derivative with respect to any vector field
\begin{equation}
l = -o^A o^{A'}
\end{equation}
tangent to the null generators of $N_0$.  In other words, $l$ is non
expanding and shear free,
\begin{equation}
\rho=\sigma=0.  
\end{equation} 

An isolated horizon $N_0$ equipped with a foliation by space-like
2-cross sections is called non rotating isolated horizon (NRIH)
whenever a transversal, future oriented null vector field
\begin{equation} 
n\ =\ -\iota^A\iota^{A'},
\end{equation}
defined on $N_0$ by  the gradient of a  function $v$ labeling the leaves 
of the foliation%
\footnote{That is, for every vector $X$ tangent to $N_0$, we have $X^a
n_{a }= X^a v_{,a}$.}
satisfies the following conditions on $N_0$: 
\medskip

i) $n$ is shear free, and its expansion is a negative function of $v$,
\begin{equation}
\lambda = 0,\ \mu = f(v)  < 0;
\end{equation}

ii) moreover, it is assumed that the Newman-Penrose spin-coefficient 
$\pi$ vanishes 
\begin{equation}
\pi = 0;
\end{equation}

iii) the Ricci tensor component $R_{\mu\nu}m^\mu\bar{m}^\nu$  
is a function, say $K$, of the function $v$ only;
\begin{equation}
R_{\mu\nu}m^\mu\bar{m}^\nu = K(v),
\end{equation}
where $m$ is a null,  complex
valued vector field tangent to the slices $v=const$ normalized by
$m^\mu {\bar m}_\mu=1$; 

$iv$) The vector field $ k^\mu = G^{\mu \nu}l_\nu$, where $G_{\mu\nu}$
 is the Einstein tensor, is causal, $k^\mu k_\mu \le 0.$
\medskip

The vanishing of the shear and of the expansion of $l$ is rescaling invariant.
We normalize $l$ such that 
\begin{equation}
l^\mu n_\mu = -1.
\end{equation}
A NRIH will be denoted by $(N_0, [(l,n)])$ where the bracket indicates, 
that the vector fields $(l,n)$ are defined up to the foliation preserving 
transformations $v \mapsto v'(v)$.  

\bigskip
 
{\bf  Space-times admitting NRIH.}\\ 
Suppose now, that $(N_0, [(l, n)])$, is a NRIH and the Einstein vacuum
equations hold in the past of a neighborhood of $N_0$.  To
characterize (locally) a general solution we need to introduce another
null surface, $N_1$ say.  Let $N_1$ be a surface generated by finite
segments of the incoming null geodesics which intersect $N_0$ at
$v=v_1$ (see (\ref{times}): we are assuming that the cartesian product
corresponds to the foliation and the variable $v$) and are parallel to
the vector field $n$ at the intersection points. Thus the
intersection,
\begin{equation}
N_0 \cap N_1 =: S 
\end{equation}
is the cross section $v=v_1$ of $N_0$. Locally (see above for the
definition of `locally'), the metric tensor is uniquely
characterized (up to diffeomorphisms)  by
Friedrich's reduced data:
\begin{eqnarray}
{\rm on}\ \ S&:& m,  {\rm Re}\rho,\ 
{\rm Re}\mu, \ \sigma,\ \lambda, \pi,\label{reducedS}\\ 
{\rm on}\ \ N_0&:&\ \  \Psi_0,\label{reduced0} \\
{\rm on}\ \ N_1&:&\ \ \Psi_4,\label{reduced1}
\end{eqnarray}
where $m$ is a complex valued vector field tangent to $S$.  The
resulting solution is given by a null frame which satisfies the
following gauge conditions,
\begin{equation}\label{gaugeM}
\nu = \gamma =\tau=\pi -\alpha-{\bar \beta}= \mu -{\bar \mu} =0,
\end{equation}
locally in the spacetime,  and
\begin{equation}\label{gaugeN}
\epsilon = 0 \ {\rm on}\ \ \ N_0. 
\end{equation}

Conversely, given submanifolds $N_0\cup N_1$ of a time oriented
4-manifold $M$, the triple $(M,N_0,N_1)$ being diffeomorphic (by the
time orientation preserving diffeomorphism) to the one above, every
freely chosen data (\ref{reducedS}, \ref{reduced0}, \ref{reduced1})
corresponds to a unique solution to the vacuum Einstein equations.
%
%
%
%

Let $(N_0, [(l,n)])$ be a NRIH . To calculate Friedrich's data, we
need to satisfy the gauge conditions (\ref{gaugeN}),\,
(\ref{gaugeM}). Since $N_0$ is non-diverging and shear-free, we can
choose on $N_0$ a normalized complex vector field $m$ tangent to the
foliation, such that $m$ is Lie constant along the null generators of
$N_0$. This implies the vanishing of $\epsilon - \bar{\epsilon}$.
From  the  generalized `0th law' \cite{mech} we
know that, if we parameterize the foliation of $N_0$ by a function
$v'$ such that
\begin{equation}
\mu' = {\rm const} \ \ {\rm on}\ \ N_0,
\end{equation}
then owing to the vacuum  Einstein's equations  
\begin{equation}
\epsilon' + \bar{\epsilon'}= {\rm const}\ \ {\rm on} \ \ N_0.
\end{equation}  
The geometric meaning of this law is that another function
\be
v\ :=\ {\rm exp}(2\epsilon v') 
\ee
defines an affine parameter along the null generators of $N_0$.
Therefore, if we use the pair $(l,n)$ corresponding to the function
$v$, then
\begin{equation}
l^\mu l_{\nu;\mu}\ =\ 0,\ {\rm hence}\ \ \epsilon = 0, \ \ {\rm on}\ \ N_0.
\end{equation}
(Incidently, in this normalization,  $l^\mu n_{\nu;\mu}\ =\ 0 $
due to $\pi = 0$).  The vanishing of $\pi -\alpha -\bar{\beta},\, \mu
-\bar{\mu}$ is automatically ensured on $N_0$ by the pullback of $n$
on $N_0$ being $dv$. Finally, the gauge conditions (\ref{gaugeM}) can
be satisfied locally in $M$ by appropriate rotations of a null frame
along the incoming geodesics, not affecting the data already fixed on
$N_0$.

Now, we can consider the reduced data of the horizon.  It follows
directly from the definition, that
\begin{equation}\label{necessary}
{\rm on}\ S:\ \sigma =\lambda = {\rm Re}(\rho)= \pi = 0, \ 
\mu = {\rm const}<0\ 
\ {\rm and \ \ on}\ N_0:\ \Psi_0\ =\ 0.\ 
\end{equation}
We also know \cite{mech} that the property (iii) in the definition of
NRIH implies that the 2-metric tensor induced on $S$ is spherically
symmetric.  The above conditions are necessary for the reduced data to
define a NRIH.

Conversely, suppose that reduced data (\ref{reducedS}),\,
(\ref{reduced0}) satisfy the conditions (\ref{necessary}) and that
they define a spherically symmetric 2-metric on the slice $S$. Then,
it follows from the Einstein vacuum equations that, locally, $N_0$ is
a NRIH.   To summarize, locally, $N_0$ is a NRIH
if and only if the vacuum space-time is given by the  reduced data
(\ref{reducedS}),\, (\ref{reduced0}),\, (\ref{reduced1}) such that
(\ref{necessary}) holds and the vector field $m$ defines on the slice
$S$ a homogeneous 2-metric tensor. The degrees of freedom are: $i)$
the radius $r_0$ of the 2-metric of $S$, and $ii)$ a complex valued
function $\Psi_4$ freely defined on $N_1$.  The constant $\mu_{|S}$
can be rescaled to be any fixed $\mu_0<0$.

\bigskip

{\bf Non existence of Killing vector fields for NRIH.}
Let us apply now our very result to the issue of the existence of
Killing vectors. The usual way one addresses that problem is writing
the Killing equation and trying to solve it.  Another way is to look
for invariant objects and see if those have a common
symmetry%
\footnote{Scalar invariants can be defined on $M$ or even on
the bundle of null directions, see Nurowski {\it et al} \cite{Nurek}}.
(Perhaps the first way is a little better to prove the existence
whereas the second way may be more useful to disprove it.)  We will
apply the second one. As it was indicated  in
\cite{phase}, a null surface admits {\it at most one} structure of
NRIH. Moreover, let as fix a number $\mu_0<0$ and use the rescaling
freedom to fix the null vector fields $(l,n)$ representing the NRIH
structure $[(l,n)]$, such that
\begin{equation}
\mu = \mu_0, \ \ {\rm on}\ \ N_0.
\end{equation}
There is exactly one pair $(l,n)$ on $N_0$ which satisfies the NRIH
properties and the normalization of $\mu$. Every isometry of spacetime
preserving $N_0$, preserves the value of $\mu$. Therefore, it
necessarily preserves the vector fields $l$ and $n$. Hence, the
potential local isometry preserves also the function
\begin{equation} 
|\Psi_4|^2\ =\ |C_{\nu\mu\alpha\beta}n^\nu m^\mu n^\alpha m^\beta|^2,    
\end{equation}
where $C$ is the Weyl tensor.  

Let us use the above isometry invariant to see whether $N_0$ admits a
tangential null Killing vector field. On $N_0$, the (would be)
Killing vector is of the form
\begin{equation}
\ksi \ =\ b_0 l
\end{equation}   
where $b_0$ is a function. The following should be true
\begin{equation}
0\ =\ b_0 l^\mu(|\Psi_4|^2)_{;\mu}\ =\ -8b_0 \epsilon |\Psi_4|^2 \ =\ 
-4 b_0 ({\rm surface\ \ gravity}) |\Psi_4|^2,
\end{equation}
the second equality being the consequence of the Einstein equations and 
the Bianchi identities. Since the surface gravity is not zero,
this contradicts the existence of a null Killing 
vector field on the horizon unless 
\begin{equation}
\Psi_4\ =\ 0.
\end{equation}

The general formula for a possible Killing vector field tangential to
$N_0$ is
\begin{equation}
\ksi\ =\ b_0 l + K
\end{equation}
where $K$ is tangent to the leaves of the foliation and together with
$b_0$ is subject to the following restrictions.  Since the isometry
generated by $\ksi$ has to preserve the foliation of $N_0$, and the
flow generated by $l$ already does, the function $b_0$ is constant on
each leaf of the foliation. Since the symmetry has to preserve the
vector field $l$, $b_0$ is constant on $N_0$ and $K$ commutes with
$l$. Finally, because the symmetry should preserve the internal
degenerate metric tensor on $N_0$, $K$ on each leaf is a Killing
vector field. On the other hand, the equation
\begin{equation}
\ksi(|\Psi_4|^2)\ = \ 0,
\end{equation} 
implies 
\begin{equation}\label{sym}
b_0\ =\ {1\over 8\epsilon} K({\rm ln}|\Psi_4|^2).
\end{equation}
For a generic $\Psi_4$ defined on the cross section $S$ of $N_0$, the
right hand side of (\ref{sym}) is not constant on $S$ for any Killing
vector field of $S$.\footnote{If it were constant, on the other hand,
then necessarily $b_0=0$ provided the orbits of $K$ in the
2-sphere are closed.} So, generically, there is no Killing vector 
field in a past neighborhood of a NRIH which is tangent to $N_0$.
(Sufficient conditions for the existence of a
Killing symmetry of an isolated horizon will be derived in a
forthcoming paper.)

\bigskip 

{\bf  General isolated horizons.}\\
A  solution admitting the general isolated horizon can also
be characterized using the reduced data. One can easily check that,
whenever
\begin{equation}\label{geniso}
\sigma\ =\ \rho\ =0, \ \ {\rm on}\ \ S, \ \ {\rm and}\ \ \Psi_0\ =\ 0,
 {\rm on}\ \ N_0, 
\end{equation} 
in the reduced data set (\ref{reducedS}), (\ref{reduced0},
(\ref{reduced1}), then the corresponding solution satisfies $\sigma =
\rho = \Psi_0 = 0$ on $N_0$, hence $N_0$ is an isolated horizon.  Of
course the above data is also necessarily an isolated horizon data.

Therefore:
\medskip

{\it $N_0$ is an isolated horizon in Einstein's vacuum space-time, if
and only if it is locally given by the reduced data (\ref{reducedS},
\ref{reduced0}, \ref{reduced1}) and the conditions (\ref{geniso}), the
remaining data ${\rm Re}\mu,\, \lambda,\, \pi$ on $S$ and $\Psi_4$ on
$N_1$ being arbitrary.}
  
\bigskip
 
{\bf The superposition method.}
There is one feature of the characteristic Cauchy problem of
\cite{Friedrich} we would like to 
emphasize more strongly here because of its relevance to
the generalization of  BH mechanics.  Given a reduced data 
(\ref{reducedS}), (\ref{reduced0}),
(\ref{reduced1}) one can evolve it, in particular, along the surface
$N_0$.  The data determines at each point of $N_0$ a vacuum solution:
a null 4-frame, the spin connection and the Weyl tensor.  Remarkably,
the evolution of every field 
along the null generators of $N_0$ is independent of $\Psi_4$
except the evolution of $\Psi_4$ itself.  We tend to think of this
construction as a non-linear superposition of a vacuum solution given
near $N_0$ with the contribution coming from data $\Delta \Psi_4$
given on $N_1$ and evolved tangentially to $N_0$.  If we know a
spacetime whose Newman-Penrose coefficients on $N_0$ we particularly
like, but $\Psi_4$ is not relevant for us, by varying $\Psi_4$ on a
transversal null surface $N_1$ we obtain a large family of solutions
each of which has the desired properties on $N_0$.  For example, let
us take the Schwarzschild metric as the preferred solution, $N_0$
being a part of its horizon. The family of solutions obtained by the
superposition with $\Psi_4$ coming in tangentially to $N_0$, is {\it
exactly the set of general vacuum solutions admitting a NRIH which we
have derived in this paper}. For every member of this family, on
$N_0$, the 4-metric tensor, and all, except $\Psi_4$, Newman-Penrose
coefficients are the same, as those of Schwarzschild. Ashtekar and
collaborators wrote the laws of BH mechanics of Schwarzshild horizon
purely in terms of the spin and curvature coefficients on $N_0$, not
involving $\Psi_4$.  That is why the laws hold automatically for a
general NRIH \cite{mech}.
 
\bigskip

{\bf Kerr like isolated horizons.}  The superposition can be well
applied to the Kerr metric. Consider reduced data given by the
following recipe:

a)  take a reduced data for Kerr, such that $N_0$ is an isolated horizon,
and $N_1$ is an arbitrary transversal null surface;

b) Keep $\Psi_0 $ on $N_0$, and ${\rm Re}\rho,{\rm
Re}\mu,\sigma,\lambda, \pi$ on the intersection $S$, but 
use an arbitrary function for $\Psi_4$ on $N_1$.

The resulting solution will be described by the same  4-metric tensor and 
the Newman-Penrose coefficients on $N_0$, except $\Psi_4$, as the original 
Kerr metric.     
  
Following this example and the definition of NRIH, we propose to
define a {\it Kerr like isolated horizon} to be a null surface $N_0$
equipped with an induced (degenerate)  intrinsic
metric tensor and the Newman-Penrose spin connection coefficients of
the Kerr solution.  This will determine the Weyl tensor spin
coefficients except $\Psi_4$.  Therefore, if we formulate the laws of
rotating BH exclusively in terms of this data on the horizon, the same
laws will hold for every metric tensor admitting the Kerr like
isolated horizon.  Since an analogous Schwarzschild like horizon would
be exactly a NRIH, the above definition is a natural step toward
defining a rotating case.

\medskip

{\bf NRIH in the Einstein-Maxwell case.} In a non-vacuum case, the
conditions imposed on a NRIH imply restrictions on the stress energy
tensor of the matter.  They are \cite{phase}
\begin{equation}
\Phi_{00}\ =\ \Phi_{01}\ =\ \Phi_{02}\ =\ 0\ =\ \delta(\Phi_{11} + 
{1\over 8}R)
\end{equation}
the last equation being the condition $iii)$ in the definition of
NRIH.  Those conditions are met by an electro magnetic field such that
\begin{equation}
\Phi_{0}=0,  
\end{equation}
and $|\Phi_1|^2$ is constant on the leaves of the foliation of $N_0$.

If we assume the Einstein-Maxwell equations to hold on $N_0\cup N_1$,
by looking at the Newman-Penrose version of the Maxwell equations, it
is easy to complete the vacuum free data with suitable data for the
electro-magnetic field. Indeed, for $\Phi_0$ given on $N_0$, $\Phi_1$
defined on $S$ and $\Phi_2$ defined on $N_1$, the Einstein-Maxwell
equations determine the metric tensor, connection, curvature and
electro magnetic field on the null surfaces $N_0$ and $N_1$, as well
as their rates of change in the transversal directions.  Then, $N_0$
is a NRIH if and only if the Einstein-Maxwell data is given by the
reduced data (\ref{reducedS}-\ref{reduced1}) of the vacuum NRIH case,
and
\begin{equation} 
\Phi_0\ = \ 0, \ {\rm on}\ \ N_0,\ |\Phi_1| \ = \ {\rm const}, \ \ 
{\rm on} S,
\end{equation}     
$\Phi_2$ being arbitrary on $ N_1$. 

The electro magnetic field affects only the evolution of the
gravitational data in the direction transversal to $N_0\cup N_1$.  In
this case the existence/uniqueness statements can be found in
Friedrich's contribution in \cite{Friedrich3}.
\bigskip

\noindent {\bf Acknowledgments.} I am most grateful to Abhay Ashtekar
for the introduction to his idea of the isolated horizons and for
numerous discussions, and to Helmut Friedrich for drawing my attention
to his characteristic Cauchy problem which was crucial for this paper.
I have also benefited from conversations with Alan Rendall, Ted
Newman, Istvan Racz, Pawel Nurowski, Bob Wald, Christopher Beetle and
Stephen Fairhurst. Finally, I would like to thank MPI for
Gravitational Physics in Potsdam-Golm and the organizers of the
workshop Strong Gravitational Fields held in Santa Barbara for their
hospitality. This research was supported in part by Albert Einstein
MPI, University of Santa Barbara, and the Polish Committee for
Scientific Research under grant no. 2 P03B 060 17.

\end{document}